# The large-scale axisymmetric magnetic topology of a very-low-mass fully-convective star

Jean-François Donati[1], Thierry Forveille[2], Andrew Collier Cameron[3], John R. Barnes[3], Xavier Delfosse[4], Moira M. Jardine[3], Jeff A. Valenti[5]

[1] *LATT, Observatoire Midi-Pyrénées,14 Avenue E. Belin, F-31400 Toulouse, France*

[2] *CFHT, 65-1238 Mamalahoa Hwy, Kamuela HI, 96743 USA*

[3] *SUPA, University of St Andrews, North Haugh, St Andrews, Fife, Scotland KY16 9SS*

[4] *LAOG, Université Joseph Fourier, 31 rue de la Piscine, F-38041 Grenoble, France*

[5] *STScI, 3700 San Martin Drive, Baltimore MD, 21218 USA*

**Understanding how cool stars produce magnetic fields within their interiors is crucial for predicting the impact of such fields, like the activity cycle of the Sun. In this respect, studying fully-convective stars enables to investigate how convective zones participate by themselves to magnetic field generation. We present here a magnetic map of a rapidly-rotating very-low-mass fully-convective dwarf obtained through tomographic imaging from time series of spectropolarimetric data. Our results, demonstrating that fully-convective stars are able to trigger axisymmetric large-scale poloidal fields without differential rotation, challenge existing theoretical models of field generation in cool stars.**

Magnetic fields of cool stars are supposedly produced and sustained through dynamo processes operating in their outer convective envelopes and involving both differential rotation and turbulence (*1*). While differential rotation winds up an initially poloidal field, so generating a toroidal field, cyclonic turbulence acting on the toroidal field regenerates a poloidal field of opposite polarity, ending the first half of the so-called “magnetic activity cycle”. In partly-convective Sun-like stars, dynamo processes presumably concentrate where differential rotation is strongest (*2*), at the interface layer between their radiative cores and convective envelopes (*3-4*). In fully-convective stars however, the lack of such layers has led theoreticians to propose that their strong magnetism may be due to dynamo processes of other kinds (*5-7*). Recent models however disagree on the type of large-scale magnetic topologies that fully-convective stars can generate; while some conclude that very-low-mass stars should produce non-axisymmetric non-cyclic fields and negligible differential rotation (*8-10*), others predict they should trigger significant differential rotation and mainly axisymmetric, oscillatory fields (*11-12*). This disagreement essentially reflects the fact that existing numerical models of stellar dynamos must rely on various simplifications that can distort their results, realistic models being still beyond the reach of high-performance computing.

Observing magnetic topologies of very-low-mass dwarfs is thus critical to progress in this field. Existing data demonstrate that fully-convective stars, and especially the rapidly rotating ones, are strongly magnetic and show both indirect and direct evidence for the presence of fields in their atmospheres (*13-16*). The broadening that fields induce in unpolarised spectral lines through the Zeeman effect (*31*) indicate that magnetic fields of several kG are present at the surface of very-low mass dwarfs (*15-16*). However, distinguishing among the various existing theoretical models requires maps of the vector magnetic field over the stellar surface. Since line-broadening measurements are poorly informative on the field topology, one solution is to analyse the polarisation signatures that magnetic fields generate in spectral lines through the Zeeman effect (*31*). Sensing both the magnetic flux and the field orientation, this complementary method has proved very successful for detecting the fields of active stars and estimating their topologies (*17-18*); it thus optimally suits the present needs.

During a 3-night observing run (on 2005 Aug. 18, 20 and 22) with the new high-efficiency high-resolution spectropolarimeter ESPaDOnS (*19,31*) recently installed on the Canada-France-Hawaii Telescope (CFHT), we obtained 64 circularly polarised spectra sampling three non-consecutive rotation cycles of V374 Peg, a rapidly rotating M4 dwarf with a low enough mass ($\approx 0.28\ M_{\odot}$, *20*) to ensure convection throughout the whole star (*21*). Using Least-Squares Deconvolution (LSD, *17,31*), we extracted the polarisation information from about 5,000 spectral lines simultaneously and produced mean Stokes $I$ (unpolarised) and $V$ (circularly polarised) profiles of V374 Peg for all collected spectra. While unpolarised LSD profiles inform about cool spots on the stellar surface through the shape distortions they produce (*22-23*), Stokes $V$ profiles sound the surface magnetic topology through the Zeeman signals it induces in spectral lines (*24-26*). Strong magnetic signatures with amplitudes of up to 0.5% of the unpolarised continuum are detected from V374 Peg; unpolarised profile distortions from cool spots are very weak.

Matching unpolarised profiles yields that the equatorial rotation velocity $v_{eq}$ of V374 Peg verifies $v_{eq}\sin i$=36.5±0.3 km/s (where $i$ notes the angle between the rotation axis and the line-of-sight) while its heliocentric radial velocity is –3.9±0.1 km/s. We find that the Zeeman signatures and their temporal variations repeat every ≈10.7 hr, a timescale consistent with the rotation period. What we observe is thus truly rotational modulation, with a period of $P_{rot}$=0.4456±0.0002 d. The dynamic circular polarisation spectrum, phased to the stellar rotation cycle (Fig.1), shows clear Zeeman signatures migrating throughout the line profile from negative to positive velocities as the corresponding magnetic regions are carried across the visible hemisphere (from the approaching limb to the receding limb) by rotation. Three such signatures can be straightforwardly identified from Fig.1, a main one crossing line centre at phase 0.5, and two weaker ones at phase 0.20 and 0.75. The line-of-sight projected (longitudinal) component of the magnetic field, computed from the first-order moment of the Zeeman signature (*17*), is always positive and peaks at 350 G (at phase 0.5). From $P_{rot}$ and $v_{eq}\sin i$, we derive that the radius $R$ of V374 Peg verifies $R\sin i$=0.32 $R_\odot$; from the absolute K magnitude (≈7.0), we obtain that $R$≤0. 37 $R_\odot$ (*27*) and thus conclude that $i$≥60°, ie the star is seen close to equator-on. As there is little probability that $i$ exactly equals 90°, we assume $i$=70° in the following (results are similar for 60°≤$i$≤80°).

Using tomographic imaging techniques (*31*), detailed surface maps of rapidly-rotating stars like V374 Peg can be reconstructed from time-resolved series of rotationally modulated profiles of spectral lines (*22-26*). Using a stellar surface imaging code based on spherical harmonics decomposition and maximum entropy principles (*28*), we reconstructed a model stellar magnetic topology reproducing the observed Stokes $V$ dynamic spectrum. The typical longitudinal resolution achieved at the equator is ≈10° or ≈0.03 rotation cycle. With the assumption that the field is potential, we obtain a synthetic Stokes $V$ dynamic spectrum (Fig.1) matching the data at noise level. No significant improvement is obtained when assuming that the field also includes a toroidal component. The recovered topology (Fig.2) features mostly radial field, with areas where magnetic strengths reach 2 kG. Azimuthal and meridional field features are typically 4 times weaker. As expected from the Zeeman signatures, featuring constant sign throughout the rotation cycle, the latitudinal polarity pattern of the reconstructed radial field map is simple, with positive fields occupying most of the upper stellar hemisphere (where the method sensitivity to radial fields is highest).

With observations of 3 rotation cycles sequentially shifted by 4.5 cycles, we can estimate the amount of differential rotation shearing the surface of V374 Peg. We assume that the rotation rate at the surface of V374 Peg depends mostly on latitude $\theta$ and varies as $\Omega_{eq}$-$d\Omega\sin^2\theta$, $\Omega_{eq}$ being the angular rotation rate at the equator and $d\Omega$ the difference in angular rotation rate between the equator and pole. We find that optimal fits to the data (*31*) are achieved for $d\Omega$ =0±0.02 rad/d (68% confidence interval), and thus that the rotation of V374 Peg is compatible with solid body rotation (Fig.3). It implies that V374 Peg needs at least 1 yr for its equator to lap the pole by one complete turn, ie more than 3 times as much as the Sun. This is compatible with recent results indicating that surface differential rotation is apparently vanishing almost completely in very-low-mass stars (*29*).

Although no measurement has been published yet for V374 Peg itself, magnetic field estimates derived from Zeeman broadening of unpolarised profiles (*31*) exist for a few similar mid M dwarfs (*16*), indicating that about half the stellar surface is covered with field strengths of 3 to 4 kG (implying overall fluxes of 1.5 to 2 kG). The average field flux we report here is about 0.5 to 1 kG, indicating that the method we use likely underestimates the actual field strengths of V374Peg. The likely explanation is that the method we use, while successful at recovering the field topology on spatial scales larger than the resolution element (about 10° in longitude at the equator), fails at detecting very small-scale multipolar magnetic groups whose average Zeeman signature (sensitive to the vector properties of the field, *31*) is zero. We thus conclude that the magnetic field of V374 Peg probably includes both a large-scale structure that we are able to detect and map, and a very-small-scale highly-multipolar structure that remains inaccessible to us. Since our aim is to compare the large-scale structure of the field with that predicted by numerical dynamo models, this limitation is not a major problem for our study.

Our image shows several striking differences with magnetic maps of non-fully-convective active stars (*18*). While magnetic topologies of partly-convective stars almost always involve a strong and dominant toroidal field component at the stellar surface without which the Stokes *V* data cannot be reproduced (*18,28*), the field of V374 Peg includes no such feature and can be modelled as a purely potential structure. As strong large-scale toroidal fields are systematically observed in the photosphere of active stars featuring (even small) differential rotation (*30*), it further strengthens our findings that V374 Peg rotates mostly as a solid body and confirms the claim that differential rotation vanishes in fully-convective stars (*29*). Another important difference is that the potential field of V374 Peg is about an order of magnitude stronger than that of partly-convective stars (*18*), with a typical strength of several kG as opposed to a few 100 G for other active stars. Its large-scale spatial structure is also much simpler, with monopolar regions covering up to one hemisphere (and thus far larger than the resolution element at the surface of the star).

The spherical harmonics expansion of the recovered magnetic field distribution (Fig.4) reveals that the dominant mode excited in V374 Peg by dynamo processes corresponds to a dipole aligned with the rotation axis (spherical harmonics degree $\ell$=1 and order $m$=0) whose amplitude reaches ≈3 kG. Power from this mode is leaking into the $\ell$=$m$=1 mode, reflecting the offset of the main radial field region with respect to the pole. Several other modes, and in particular sectorial modes ($\ell$=$m$), are also detected on V374 Peg; the $\ell$=$m$=1 and $\ell$=$m$=3 modes are clearly visible on the reconstructed azimuthal and radial field maps respectively (Fig.2). All are however at least 2.5 times weaker than the main axisymmetric mode seen on V374 Peg. Little power is detected for modes with degrees larger than 5 despite our ability to recover modes with up to $\ell$=18, emphasising again that the magnetic features detected on V374 Peg are far larger than the resolution element.

This result challenges the most recent theoretical models of fully-convective stars (*8-12*). On one side, models predicting weak surface differential rotation, in agreement with observations, conclude that dynamo processes should only produce non-axisymmetric large-scale fields (*8-10*), in contradiction with our result. On the other side, dynamo models that succeed at producing large-scale axisymmetric fields, in rough qualitative agreement with our findings, predict significant surface differential rotation (*11-12*), in contradiction with our observations. Moreover, such models also predict that the toroidal component of the dynamo field should dominate, even at photospheric level, which again is not supported by our observations. We thus conclude that existing numerical dynamo models, even though rather elaborated, are not yet close enough to reality to predict successfully the magnetic topologies of convective stars; spectropolarimetric observations of a large sample of cool stars with ESPaDOnS should provide the strong observational constraints that models require .

32. **Acknowledgements** This paper is based on observations obtained at the Canada-France-Hawaii Telescope (CFHT) which is operated by the National Research Council of Canada, the Institut National des Science de l'Univers of the Centre National de la Recherche Scientifique of France, and the University of Hawaii.

**Supporting Online Material**

www.sciencemag.org

Materials and Methods

Figs. S1, S2

**Figure 1**: Dynamic Stokes *V* spectrum of V374Peg phased to a rotation period of 0.4456 d. **a**: LSD profiles derived from observations. **b**: Fit to the data obtained with our magnetic imaging code. The vertical lines depict the blue edge, centre and red edge of the rotationally broadened line profile of V374 Peg (in the star velocity rest frame). Zeeman signatures migrate throughout the profile from negative to positive velocities (thus producing slanted tracks in the dynamic spectrum) as parent magnetic regions are carried across the visible hemisphere (from the approaching limb to the receding limb) by rotation. Black and white correspond to relative circular polarisation levels of –0.3 and 0.3%.

**Figure 2**: The mostly radial magnetic topology at the surface of V374Peg, as reconstructed from our observations. **a**: Radial field component. **b**: Azimuthal field component. **c**: Meridional field component. The star is shown in flattened polar projection down to latitudes of –30°, with the equator depicted as a bold circle and parallels as dashed circles. Phase coordinates are noted around each plot, with external radial ticks indicating observations. Black and white correspond to field strengths of –1.8 and 1.8 kG.

**Figure 3**: Variation of statistical $\chi^2$ with differential rotation parameters $\Omega_{eq}$ and $d\Omega$. The total number of fitted data points is ≈3,500, implying that the grey ellipse (enclosing points with $\Delta\chi^2 \leq 9$) projects on each axis into confidence intervals of 99.7%. Rotation of V374 Peg is compatible with solid body rotation at a rate of 14.100±0.004 rad/d. Black and white correspond to $\chi^2$ values of 3555.3 and 3564.3 ($\Delta\chi^2$=0 and 9).

**Figure 4**: Mode amplitude in the spherical harmonics expansion of the reconstructed magnetic field distribution, as a function of mode degree $\ell$ and order *m*. **a**: Radial field component. **b**: Azimuthal field component. **c**: Meridional field component. The strongest mode corresponds to a dipole aligned with the rotation axis, with an amplitude of ≈3 kG. Modes with orders and degrees lower than 5 only are shown here. White and black corresponds to mode amplitudes of 0 and 2.5 kG.

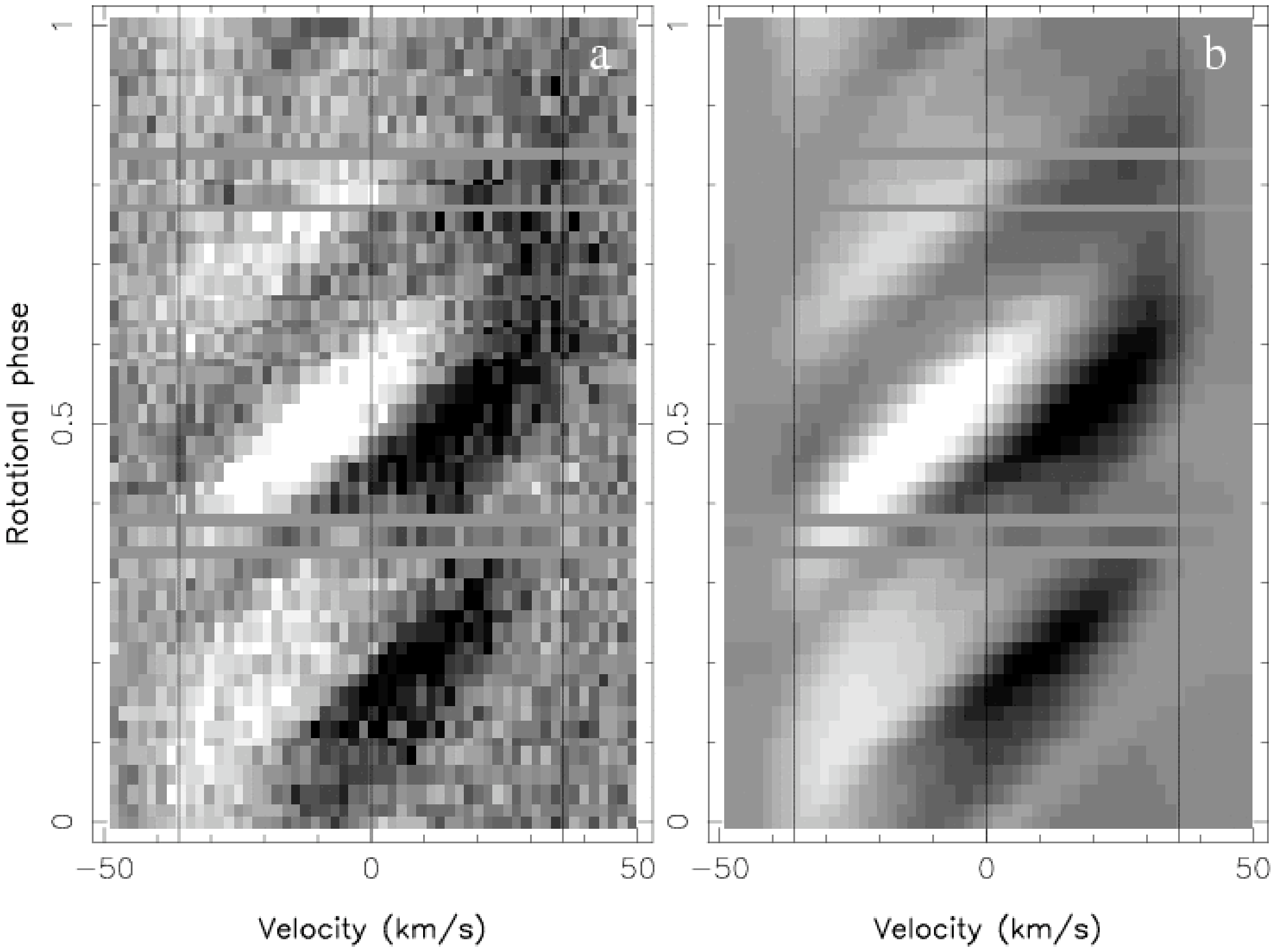

**Figure 1**

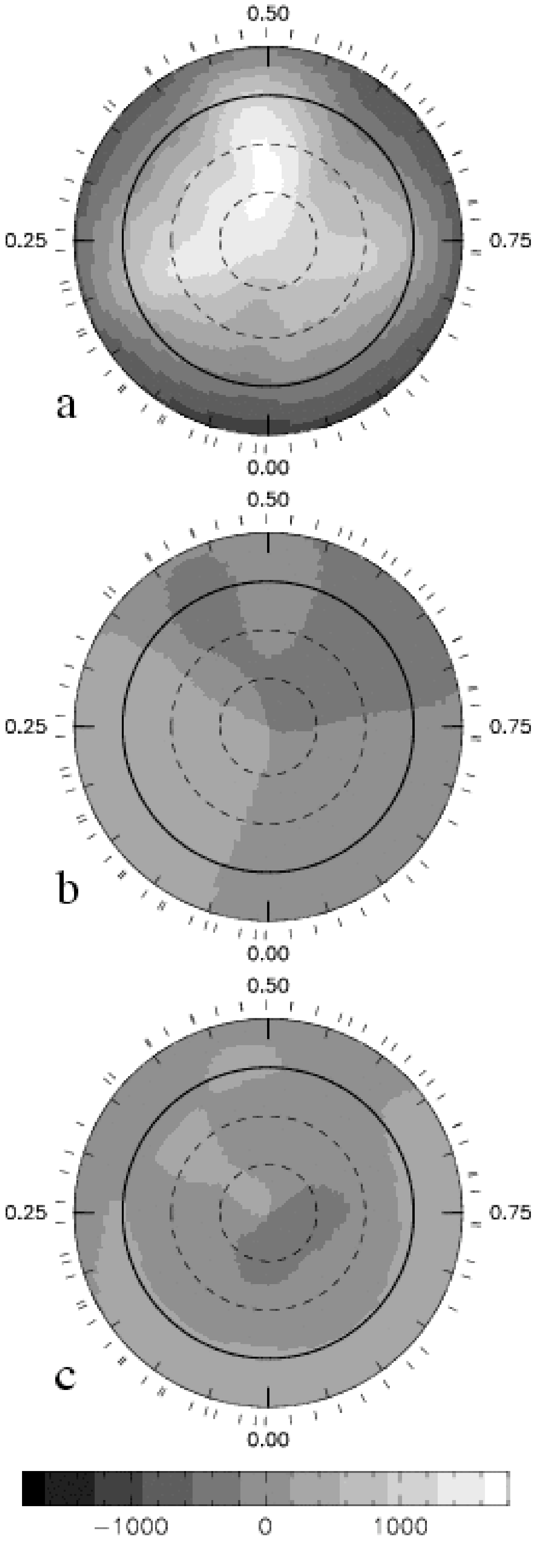

**Figure 2**

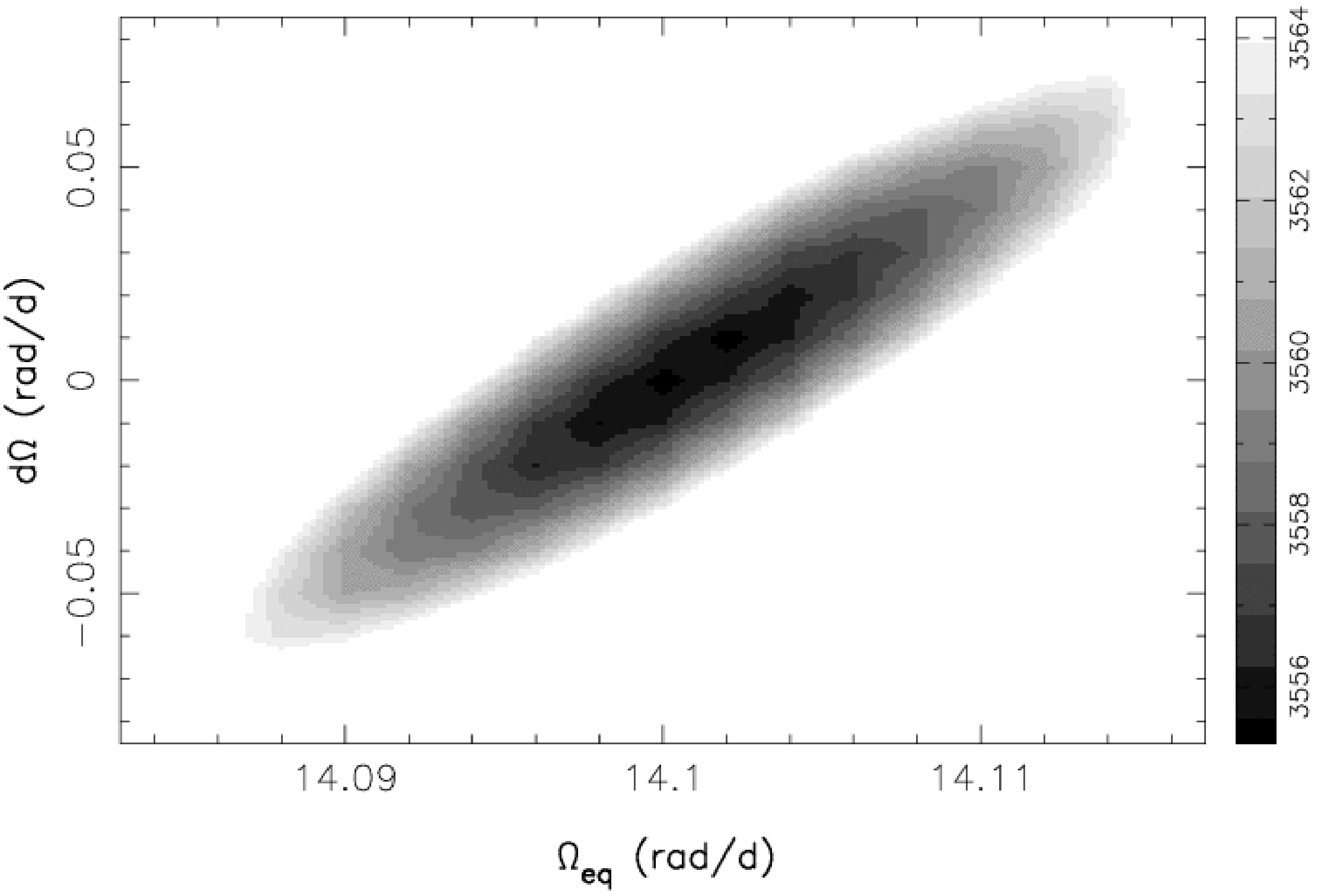

**Figure 3**

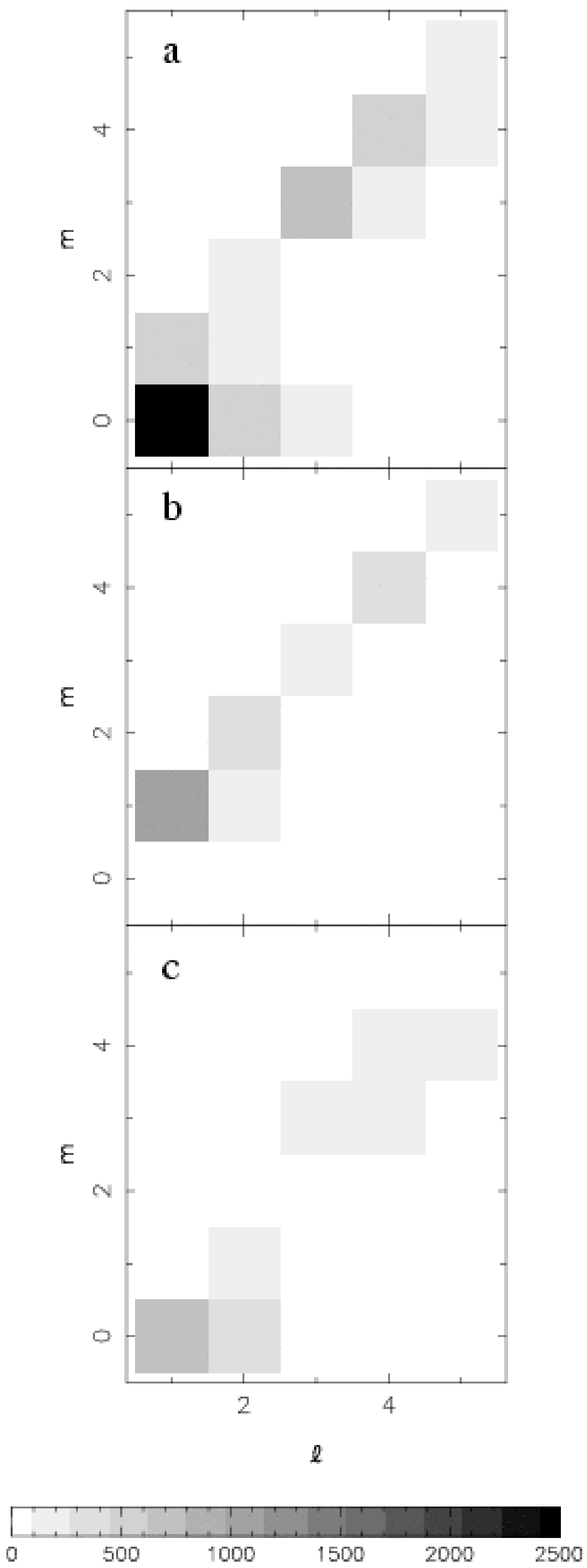

**Figure 4**

## Supporting Online Material

### Methods.

**Detecting stellar magnetic fields using the Zeeman effect.** The most successful technique to detect magnetic fields in stellar atmospheres involves measuring the Zeeman effect in atomic lines of stellar spectra. In the presence of magnetic fields, atomic spectral lines are observed to split into multiple subcomponents: this is the Zeeman effect, named after Pieter Zeeman, who discovered it in 1897. These subcomponents can be classified in 3 groups; one of them (the π components) is symmetric with respect to the line centre, while the two others (the σ components) are respectively shifted towards each side of the line centre, by an amount $\Delta\lambda$ that depends on the intensity of the magnetic field $B$, on the magnetic line sensitivity $g$ (called Landé factor) and on the line centre wavelength $\lambda$, according to $\Delta\lambda = 4.67\ 10^{-12}\ \lambda^2\ g\ B$ (with $\lambda$ and $\Delta\lambda$ in nm, and $B$ in G). Moreover, the π components are linearly polarised while the σ component are circularly polarised. In unpolarised light (Stokes I), the effect of the magnetic field is thus to broaden spectral lines, and potentially even to split them into individual subcomponents if the magnetic field is strong enough to drive the π and σ components out of their mutual overlap regime. Measuring this broadening (or splitting) yields estimates of the fraction of the stellar surface covered with fields and of the magnetic strength within this region. Little information can be extracted on the field orientation and topology, the advantage being that regions of opposite polarities can both contribute to the detected signal. In polarised light, the effect of magnetic fields is to induce circular and linear polarisation signatures within line profiles, whose strengths depends both on the field flux and its orientation with respect to the line-of-sight. At first order, circular polarisation signatures are proportional to the line-of-sight projected (longitudinal) component of the vector magnetic field. Circular polarisation (Stokes V) signatures from regions of similar flux and opposite polarities will thus cancel out mutually and produce a null global signature, if associated to similar Doppler shifts; if these regions are associated to different Doppler shifts (thanks to stellar rotation for instance, see Supporting Online Method #4), circular polarisation signatures from regions of opposite polarities no longer cancel out and can be both detected.

**Spectropolarimetry with ESPaDOnS.** The instrument includes an achromatic polarimeter, featuring rotatable retarders and installed at the Cassegrain focus of the 3.6m Canada-France-Hawaii Telescope atop Mauna Kea. It fibre feeds a bench-mounted high-resolution spectrograph, yielding full coverage of the 370 to 1,000 nm wavelength range in a single exposure. For this run, polarimetric exposures were split in sequences of 4 consecutive 300 s subexposures collected at different orientations of the polarimeter retarders, each subexposure including both orthogonal states of the selected polarisation. This procedure allows to suppress all spurious polarisation signals at first order. The full efficiency of the instrument is about 12%, yielding peak signal to noise ratio (S/N) of about 180 per 2.6 km/s velocity bin. Wavelength calibrated unpolarised and polarised spectra corresponding to each observing sequences are extracted with the dedicated software package Libre-ESpRIT, following the principles of optimal extraction.

**Deriving mean line profiles with Least-Squares Deconvolution.** LSD is a cross-correlation type technique that allows to obtain average unpolarised and polarised line profiles with largely enhanced signal to noise ratio, using simultaneously thousands of spectral features forming roughly in the same disc region. Line lists are derived from spectrum synthesis through model atmospheres. For this study, we mainly used a line list corresponding to an early M spectral type matching that of V374 Peg. About 5,000 intermediate to strong atomic spectral lines are used simultaneously to retrieve the average polarisation information in line profiles, with typical noise levels of ≈0.06% (relative to the unpolarised continuum level) per 1.8 km/s velocity bin and per individual polarisation spectrum, corresponding to a multiplex gain in S/N of about 10 with respect to a single average line analysis.

**Mapping the surface magnetic topology.** In rapidly rotating stars, the Doppler effect induces a strong correlation between the spatial location of magnetic features at the surface of the star and the Zeeman signatures they generate in line profile (Fig.S1). LSD Stokes *V* signatures of such stars appear as one-dimension magnetic maps, resolved in the Doppler direction but smeared out along isovelocity strips. Rotational modulation of Stokes *V* signatures helps removing this ambiguity and enables to extract information about the field orientation. Equatorial features produce Zeeman signatures visible for only half a rotation cycle and migrating through the full width of the line profile, and can be distinguished from polar features, visible for almost the full rotation cycle and affecting the profile core only. Similarly, azimuthal field features generate Zeeman signatures switching sign at mid transit through the line profile, and can be recognised from radial field features producing Zeeman signatures with constant sign throughout their migration (Fig.S2). Time-resolved series of circularly polarised profiles densely sampling the rotational cycle can thus be inverted into a surface map of the parent magnetic topology. To perform the numerical inversion, we use a tomographic imaging code based on principles of maximum-entropy image reconstruction; the algorithm looks iteratively for the simplest field topology (described through its spherical harmonics expansion) that matches the observed Zeeman signatures down to noise level. Given the equatorial resolution of ≈10° (implying 36 resolved element around the stellar equator), we can truncate the spherical harmonics expansion at order ≈18 with no loss of spatial information. With this method, we can search for specific field configurations, like potential, linear force-free or toroidal fields and linear combinations thereof, by imposing constraints between the spherical harmonics coefficients of the various field components.

**Estimating surface differential rotation.** We assume in the imaging process that the surface distribution is sheared by differential rotation, and that the rotation rate at the surface of V374 Peg depends mostly on latitude $\theta$ and varies as $\Omega_{\rm eq}-d\Omega\sin^2\theta$, $\Omega_{\rm eq}$ being the angular rotation rate at the equator and $d\Omega$ the difference in angular rotation rate between the equator and pole. To obtain an estimate of the surface shear, we reconstruct, for many different values of $\Omega_{\rm eq}$ and $d\Omega$, magnetic images that all contain the same amount of information. For each reconstructed magnetic image, we compute the statistical $\chi^2$ describing how well the synthetic and observed circular polarisation profiles match, using the standard formula $\chi^2 = \sum_i (O_i-S_i)^2/\sigma_i^2$ where $O_i$, $S_i$ and $\sigma_i$ respectively note the $i^{\rm th}$ observed data point, synthetic data point and corresponding error bar. The surface describing the $\chi^2$ fit to the data as a function of $\Omega_{\rm eq}$ and $d\Omega$ features a minimum;  fitting a paraboloid to this surface around the minimum gives access to the accurate coordinates of the minimum and to the surface curvature at this point, yielding optimal estimates of $\Omega_{eq}$ and $d\Omega$ and associated error bars.

**Figures.**

**S1: Basic principles of Doppler imaging.** Rotation induces a strong correlation between the spatial location of features at the surface of a star, and their signatures in spectral lines. In this example, we consider a model star with the same rotation rate and inclination angle as V374 Peg (upper panel); the star, shown at rotational phase 0.15, features a 2 kG radial field magnetic region centred at phase 0.3. At rotational phase 0.15, the magnetic region appears on the approaching stellar limb and produces a line profile circular polarisation (Stokes V) signature shifted to negative velocities with respect to line centre (lower panel), ie at a spectral position reflecting the spatial location of the parent magnetic region.

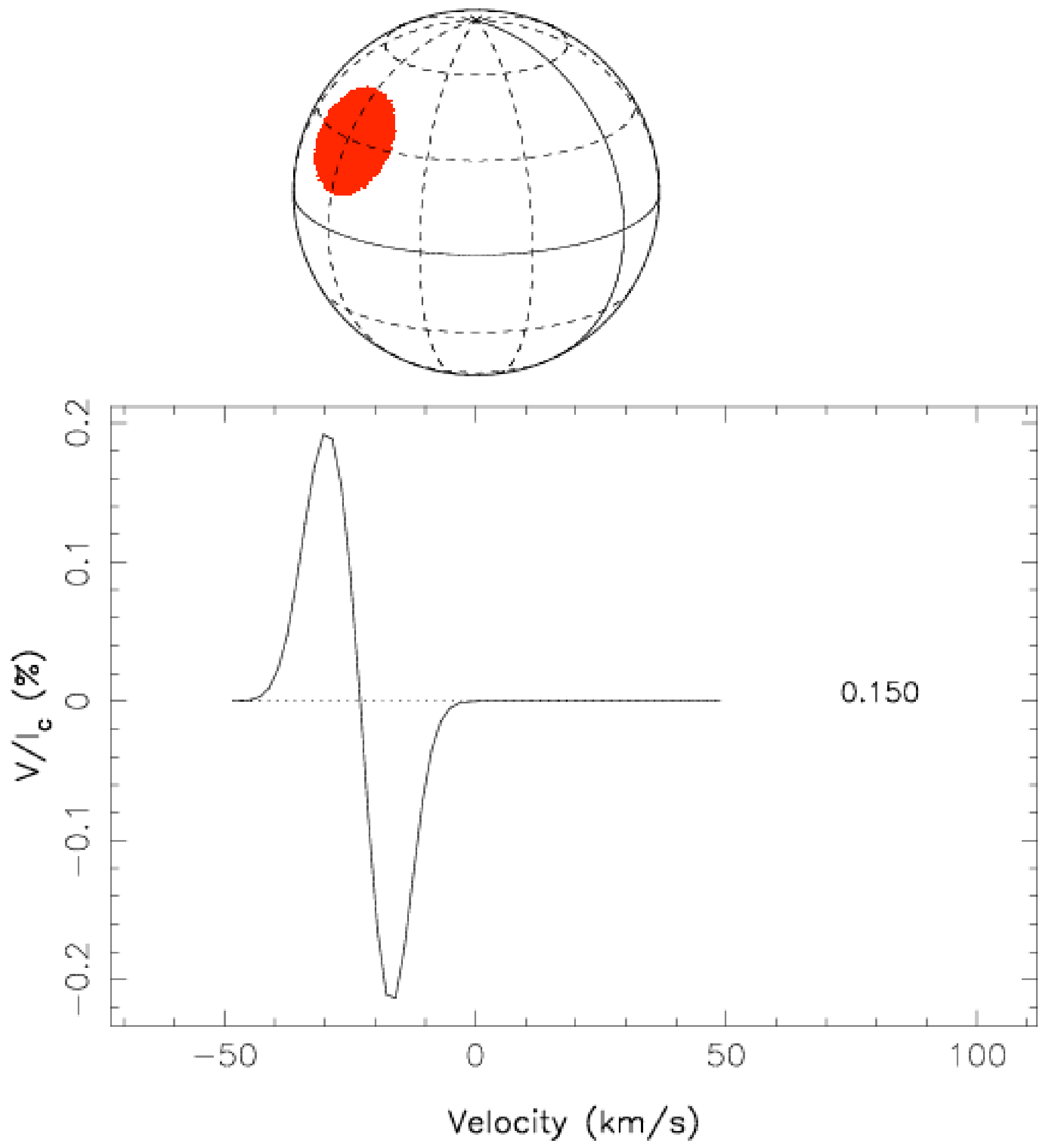

**S2: Sentivity of Zeeman-Doppler Imaging to magnetic field oriention.** This dynamic spectrum illustrates how circular plarisation signatures from magnetic spots are modulated by rotation, depending on field orientation. In this example, the star hosts two 2 kG magnetic regions at latitude 30°, one with radial field (at phase 0.3) and one with azimuthal field (at phase 0.7). The signature from the radial field region (first half rotation cycle) keeps a constant sign throughout the line profile, while that from the azimuthal field spot (second half rotation cycle) switches sign at mid transit. Red and blue correspond to relative circular polarisation levels of –0.3 and 0.3% respectively.

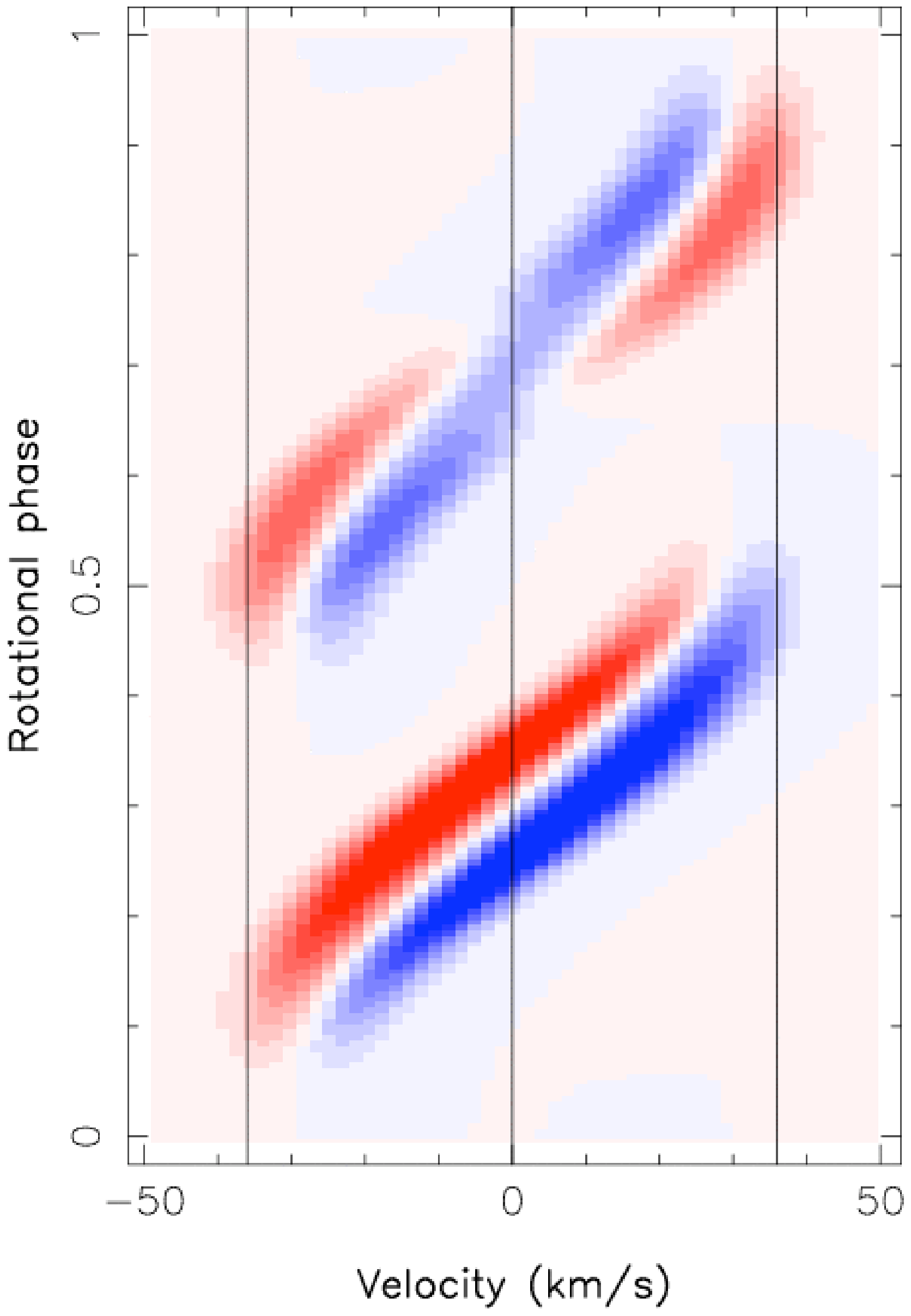